\documentstyle[aps,prl,epsf]{revtex}
\begin{document}
\draft \twocolumn[\hsize\textwidth\columnwidth\hsize\csname
@twocolumnfalse\endcsname 
\title{Critical dynamics of singlet and triplet excitations in
strongly frustrated spin systems}
\author{O. P. Sushkov, J. Oitmaa,  and
 Zheng Weihong} \address{School of Physics, University of New South
 Wales, Sydney 2052, Australia}

%\date{\today}
\maketitle
\begin{abstract}
%Various new susceptibilities to probe spin singlet excitations
%for the two-dimensional $J_1-J_2$  model have been computed
%using  Ising series expansions, and the results show that the

New data for the two-dimensional $J_1-J_2$ - model shows that the
critical
points for singlet and triplet excitations near $J_2/J_1\approx 0.38$ are 
very close together, or coincident. We propose that this is a more general
result: in strongly frustrated spin systems the critical dynamics of
singlet and triplet excitations are closely related. An effective field
theory for such a critical point, or points, is developed, and its
implications discussed. The analysis suggests the existence of gapless
or almost gapless singlet excitations in the magnetically disordered
phase.   
\end{abstract}

 ]

Quantum phase transitions in quantum 
antiferromagnets have attracted a great deal of attention
during the last decade. This field contains numerous previously 
unknown phenomena. Investigation of quantum phase transitions
in quantum antiferromagnets is very important for quantum magnetism, but
is of even wider importance because it allows one to construct and to 
analyze effective nonperturbative field theories that can be used 
in other parts of physics. 
In this paper we concentrate on rotationally invariant (2+1)D systems.
Transitions described by the nonlinear O(N) field
theory are very well understood, see, e.g., Refs. \cite{ZZ,GZ,Sach}
The most widely known example is the O(3) theory that describes a real
vector field ${\bf v}$ with Lagrangian density
\begin{equation}
\label{lv}
{\cal L}_v={1\over{2}}\left[(\partial_{t}{\bf v})^2-c^2_v(\nabla{\bf v})^2
-m^2_v{\bf v}^2\right]-{{u_v}\over{4}}\left({\bf v}^2\right)^2.
\end{equation}
The effective mass vanishes at some critical point, 
$m_v^2=a_v(g-g_{cv})$, where $g$ is some generalized ``coupling constant'',
and $a_v$ is some coefficient.
It has been argued by Haldane \cite{Haldane} and Chakravarty, Halperin and
Nelson \cite{Chakravarty} that this theory describes a quantum phase
transition between the N\'eel state ($g < g_{cv}$)
and the magnetically disordered state ($g > g_{cv}$)
in a two-dimensional quantum antiferromagnet. 
Perhaps the simplest concrete example of such a system is
the Heisenberg antiferromagnet on a double layer square lattice.
In this case  
%the  field ${\bf v}$ coincides with the spin dimer triplet excitation, and 
the ``coupling constant'' is equal to the ratio of interlayer and
inlayer antiferromagnetic interactions, $g=J_{\perp}/J$, 
$g_{cv} \approx 2.525$, see Ref. \cite{Hida}.
For $g > g_{cv}$ the excitation spectrum of the system (\ref{lv}) is gapped,
and for $g < g_{cv}$ the spectrum is gapless because of Goldstone
spin waves.

The O(1) field theory describes a real scalar field $s$ with Lagrangian
\begin{equation}
\label{ls}
{\cal L}_s={1\over{2}}\left[(\partial_{t}s)^2-c^2_s(\nabla s)^2
-m^2_s{s}^2\right]-{{u_s}\over{4}}s^4,
\end{equation}
where $m_s^2=a_s (g_{cs}-g)$. The field expectation value $<s>$ is
zero at $g < g_{cs}$ and nonzero at $g > g_{cs}$. The spectrum is gapped
in both cases. A simple example of such a system is a Heisenberg 
antiferromagnet on a "planar pyrochlore" lattice where the entire low energy
dynamics is driven by spin singlet excitations \cite{KZS}.

The situation in strongly frustrated spin systems is much more complicated
because the dynamics of triplet and singlet excitations are not
independent. The 2D $J_1-J_2$ model is an ideal testing ground for
understanding of such dynamics.
The Hamiltonian of the model reads:
\begin{equation}
\label{ham}
H = J_{1} \sum_{nn}{\bf S}_{i}\cdot {\bf S}_{j} + J_{2} \sum_{nnn} {\bf
  S}_{i}\cdot {\bf S}_{j},
\end{equation}
where $J_{1}$ is the nearest-neighbor, and $J_{2}$ is the frustrating
next-nearest-neighbor Heisenberg exchange on a square lattice.  
Both couplings are antiferromagnetic, i.e.  $J_{1,2}>0$ and
the spins $S_{i}=1/2$. 
At small $J_2$ the system exhibits N\'eel ordering and at
small $J_1$ it exhibits  collinear ordering. There is
also a general consensus on the disappearance of the magnetic ordering
at $0.38 < J_2/J_1 < 0.6$, see, e.g., Ref.\cite{Kotov}.
The structure of the magnetically disordered state is of great interest.
An important insight into the disordered regime was achieved by
 field-theory methods \cite{Read,Kotov} and dimer series expansions
 \cite{series,series2,Kotov}.
  The above works have  shown that the ground
 state in this regime is dominated by short-range singlet
 (dimer) formation in a columnar pattern. 
%The stability of such a configuration implies that the lattice symmetry is
% spontaneously broken and the ground state is four-fold degenerate. 
More recent studies based on the  Green function Monte Carlo method
favored either a plaquette state \cite{Capriotti} or a dimer state
with plaquette modulation \cite{Jongh}.
And finally the same group that claimed stability of the plaquette
state \cite{Capriotti}, now claims stability of the translationally
invariant RVB state \cite{Capriotti1}.
The above series expansions \cite{series,series2,Kotov},
as well as Monte Carlo numerical calculations
\cite{Capriotti,Jongh,Capriotti1} in various forms, have difficulties in
distinguishing between states with very close energies, as occurs in the
$J_1-J_2$ model.
In such approaches one starts from some
trial wave function (dimer, plaquette, RVB,...) and such an ansatz 
is always very restrictive:
it is impossible to get a translationally invariant state using
the dimer or plaquette expansions, and on the other hand it is impossible
to get a spontaneous violation of the translational symmetry starting
from the trial RVB state that is explicitly translationally invariant.

A novel approach was implicitly formulated in our work \cite{Sus},
see also Ref. \cite{SK}.
The idea of the approach is to use numerical calculations only in the
well established N\'eel phase at $J_2/J_1 < 0.38$. This calculation allow
us 
to find all the relevant quasiparticles in the system, and hence to determine 
the effective field theory describing the dynamics of the system. Since the 
transition is of the second order the same field theory describes the 
magnetically disordered phase and we have no problem to describe this phase.
In the present work we explicitly formulate the approach, present new,
much more accurate, numerical data, derive the effective Lagrangian of
the model, and finally discuss implications of this effective Lagrangian.
%In the N\'eel state one can use any numerical method that gives both the
%ground state and the spectrum of excitations. In this work we use the 
%series expansions method.

For the $J_1-J_2$ model we define the generalized coupling constant as
$g=J_2/J_1$. It is well established that at $g < g_{cv}$ there is
N\'eel ordering in the system. 
There are 3 independent estimates for $g_{cv}$. The first comes from 
dimer series expansions, $g_{cv}\approx 0.38$, see Ref. \cite{Kotov}.
However this estimate is problematic since it already assumes  
spin dimer order at $g > g_{cv}$. The second estimate, $g_{cv}= 0.4 \pm 0.02$,
comes from Ising series expansions for the staggered magnetization in
the N\'eel phase \cite{series2}. The third estimate $g_{cv}= 0.38 \pm 0.03$
comes from the Ising series expansions for spin-spin correlators in
the N\'eel phase \cite{Sus}. Using the second and the third estimates, and 
averaging them as independent ones we find
\begin{equation}
\label{gcv}
g_{cv}=0.39 \pm 0.016 \to 0.39 \pm 0.02.
\end{equation}
 The N\'eel ordering corresponds
to the condensation of a vector field, therefore there is no doubt
that some effective Lagrangian determines the dynamics of the field 
both at $g< g_{cv}$ and $g > g_{cv}$. 

However the vector excitation is not the only relevant
degree of freedom in the system. There are also the low energy singlet 
excitations.
 We do not have a  reliable numerical technique for direct calculation of 
singlet excitations, but we do have a well developed series 
expansion technique for  calculation of static susceptibilities.
A static susceptibility is proportional to the corresponding Green's function 
at zero frequency
$\chi_{\bf q} \propto G_{\bf q}(\omega=0) \sim Z_{\bf q}/\omega_{\bf q}^2$,
where $\omega_{\bf q}$ is the quasiparticle energy, and $Z_{\bf q}$ is the
quasiparticle residue.
So at the critical point  $1/\chi$ must vanish approximately as 
$(g-g_c)^{\gamma}$, with $\gamma = \nu(2-\eta)$, where $\nu$ is the critical 
index for  the spectral gap, $\Delta \propto (g-g_c)^{\nu}$, and $\nu\eta$ is
the critical index for the quasiparticle residue, 
$Z \propto (g-g_c)^{\nu\eta}$. For the simple O(1) Lagrangian (\ref{ls})
the critical index is $\gamma \approx 1.2$ (see, e.g., Ref. \cite{GZ}).

In this paper we use Ising series expansion\cite{series2} techniques 
to compute the susceptibility of this system.
The linked-cluster expansion method has been 
reviewed in a recent article\cite{gel00}, and will not be repeated
here. 

In Ref. \cite{Sus} we have calculated the susceptibility of the 
N\'eel state with respect to the external  field.
\begin{equation}
\label{FD}
F_d=f\sum_{i,j}(-1)^{i} {\bf S}_{i,j}\cdot {\bf S}_{i+1,j},
\end{equation}
where the coefficient $f$ gives the strength of the field.
The susceptibility is defined in the usual way
\begin{equation}
\label{chi1}
\chi=\left.{1\over{2}}{{\partial^2 E}\over{\partial^2f}}\right|_{f=0},
\end{equation}
where $E$ is the ground state energy.
This susceptibility probes spin singlet excitations that correspond to
a spin columnar dimerization.
%In this calculation we used the usual Ising series expansion up to seventh
%order. 
The susceptibility is diverging at $g=J_2/J_1 \sim 0.4$ and this is
a clear indication of the critical singlet excitation.

Unfortunately convergence of the series expansion for $\chi$ is
somewhat erratic and therefore the error bars in the numerical 
data \cite{Sus} are rather large. 
In the present work we calculate susceptibilities $\chi_z$ and
$\chi_{\perp}$ with respect to the fields
\begin{eqnarray}
\label{FDz}
&&F_{dz}=f\sum_{i,j}(-1)^{i}  S_{i,j}^{(z)} S_{i+1,j}^{(z)},\\
&&F_{d\perp}=f\sum_{i,j}(-1)^{i} \left(S_{i,j}^{(x)} S_{i+1,j}^{(x)}
+S_{i,j}^{(y)} S_{i+1,j}^{(y)}\right).\nonumber
\end{eqnarray}
Here $z$ is direction of the staggered magnetization in the
N\'eel state. Both $\chi_z$ and $\chi_{\perp}$ probe the same
excitations as $\chi$. 
%In the calculation of $\chi_z$ and $\chi_{\perp}$
%we use the usual Ising series expansion up to seventh order.
Using the Ising expansion about the N\'eel order\cite{series2}, 
we have computed the series for  $\chi_z$ and $\chi_{\perp}$
in powers of exchange anisotropy to order 7 for given 
values of $J_2/J_1$.
The calculations involve a list of 28474 distinct
connected clusters up to 9 sites.
We do not display the series
here but can provide them on request.
Fortunately convergence of the series for $\chi_{\perp}$ and $\chi_z$
 is much better than that for
$\chi$. Let us note that the field $F_d$ can be represented as
$F_d=F_{dz}+F_{d\perp}$. However the susceptibility $\chi$ cannot
be represented as a simple superposition of $\chi_z$ and $\chi_{\perp}$,
$\chi \ne \chi_z + \chi_{\perp}$. Therefore a relatively poor convergence
of the series for $\chi$ does not contradict the better convergence
for $\chi_z$ and $\chi_{\perp}$.
The results for $1/\chi_z$ and $1/\chi_{\perp}$ are presented
in Fig.1 and Fig.2. 
The susceptibilities are diverging at the critical
point $g_{cs}$. From Fig.1 we conclude that $g_{cs}=0.38\pm 0.04$, and
from Fig.2 we conclude that $g_{cs}=0.41\pm 0.03$. Considering these
estimates as independent ones and averaging them we find
\begin{equation}
\label{gcs}
g_{cs}=0.40 \pm 0.024 \to 0.40 \pm 0.02.
\end{equation}
We have also calculated the susceptibility $\chi_p$ with respect to the field
\begin{equation}
\label{Fp}
F_{p}=\sum_{i,j}(-1)^{i+j} \left(S_{i,j}^{(x)} S_{i+1,j+1}^{(x)}
+S_{i,j}^{(y)} S_{i+1,j+1}^{(y)}\right),
\end{equation}
that probes  excitations of plaquette symmetry, but this is not 
coupled to the
spin-dimer excitations. 
The Ising series for $\chi_p$ has been computed up to order 7.
The calculations involve a  list of connected clusters different
from that for $\chi_z$ and $\chi_{\perp}$ because
the symmetry of the field is different.
The number of cluster in this calculation is 28450, also up to 9 sites.
The results of the Ising series expansion
calculations for $1/\chi_p$ are shown in Fig.3. According to the data
the plaquette excitation is not critical and hence is not relevant.

The susceptibilities (\ref{FDz}) probe the real scalar field $s$
that corresponds to the spin columnar dimerization.
The ground state expectation value of the field is zero at $g< g_{cs}$ 
(no spontaneous dimerization), but
the susceptibilities are diverging at $g_{cs}$.
Therefore one can say the effective Lagrangian for the
field $s$ is given by Eq. (\ref{ls}). This is a possible scenario, but
does not explain why the critical points for the scalar
field (\ref{gcs}) and for the vector field (\ref{gcv}) coincide or at
least are very close: in this scenario this is just a pure coincidence.
To be absolutely correct we have to say that there is another scalar
field $s'$ that corresponds to the replacement $i\to j$, $j \to i$
in Eqs. (\ref{FDz}). So the symmetry of the transition is $O(1)\times O(1)$.
However this additional $s'$-fiels is completely identical to $s$, and
therefore it does not require a special consideration.

The fields ${\bf v}$, $s$ and $s'$ are not independent:
they certainly interact with each other. Therefore there is an
alternative scenario that naturally explains the coincidence of the quantum
critical points. Consider the effective Lagrangian of an interacting massive
vector field and two massless scalar fields.
\begin{eqnarray}
\label{lvs}
{\cal L}&=&{1\over{2}}\left[(\partial_{t}{\bf v})^2-c^2_v(\nabla{\bf v})^2
-m^2_v{\bf v}^2\right]-{{u_v}\over{4}}\left({\bf v}^2\right)^2\\
&+&{1\over{2}}\left[(\partial_{t}s)^2-c^2_s(\nabla s)^2+
(\partial_{t}s')^2-c^2_s(\nabla s')^2\right]\nonumber\\
&-&{{u_s}\over{4}}(s^2+s'^2)^2-{{u'_{s}}\over{4}}s^2s'^2
-{{u_{sv}}\over{4}}{\bf v}^2(s^2+s'^2),\nonumber
\end{eqnarray}
where $m_v^2=a_v(g-g_{cv})$. This Lagrangian incorporates all
the relevant degrees of freedom.
At $g< g_{cv}$ the vector field develops the
ground state expectation value 
$ <{\bf v}>=\sqrt{a_v(g_{cv}-g)/u_v}$.
and, hence, because of the interaction the massless scalar fields attain
an effective mass $m_{eff}^2=a_v(g_{cv}-g)u_{sv}/(2u_v)$.
Thus in this scenario the scalar susceptibilities are diverging exactly
at the same point where the N\'eel order disappears, 
$\chi \propto 1/(g_{cv}-g)$.
In the magnetically disordered phase at $g > g_{cv}$
both the scalar fields remain massless. This scenario corresponds to a
translationally invariant spin liquid state.
One can also introduce in Eq. (\ref{lvs}) a small negative bare mass for 
the scalar fields, $m_s^2=-|m_s^2|$. However due to Eqs. (\ref{gcv}) and
(\ref{gcs}) this mass must be very small because it splits the critical
points. In this case there is a region between the critical points where
the spin-dimer and the N\'eel order coexist \cite{Sus,SK}.

We mention, in this connection, that in one-dimensional systems the 
Lieb-Schultz-Mattis (LSM) theorem guarantees that a uniform phase 
in a quantum spin system is gapless and a gapped phase breaks translational 
symmetry \cite{LSM}. Extension of the LSM theorem to two dimensions was
proposed \cite{Af} but not proven (see also Ref. \cite{ML}).
We note that the effective Lagrangian (\ref{lvs}) 
%deduced from the numerical data 
is in  perfect agreement with the conjectured
two-dimensional  extension of the LSM theorem.
If the bare singlet mass $m_s^2$ is zero then the magnetically disordered
state is uniform and there are two gapless excitations in the state.
If $m_s^2=-|m_s^2|\ne 0$ then one of the scalar fields develops the
ground state expectation value $<s>=\sqrt{|m_s^2|/u_s}$
(spontaneous dimerization) and another attains finite positive mass due to 
the interaction terms, $m_{eff}^2={{u_s'}\over{2u_s}}|m_s^2|$.

The effective Lagrangian (\ref{lvs}) corresponds to the translationally
invariant intermediate spin liquid state (SL) and in this part it is 
in agreement with
the conclusion of the recent work \cite{Capriotti1}. However, as we have
shown, this SL state must have gapless singlet excitations.
Therefore: 1) the dimer-dimer correlation function must fall off as
 $\propto 1/r$, where $r$ is the separation between dimers, 
2) for a finite cluster of $N$ sites the static susceptibility 
 with respect to the dimerization field (\ref{FD}) must scale  as
\begin{equation}
\label{chisl}
\chi ={1\over{2f^2}} {{|<0|F_d|s>|^2}\over{E_0-E_s}} \propto
{1\over{f^2}}{{(f\sqrt{N})^2}\over{\left(1/\sqrt{N}\right)}} 
\propto N^{3/2}.
\end{equation}
Both these conclusions disagree with results of Ref. \cite{Capriotti1}.
In our opinion this can be related to the variational nature of that
calculation. We would like to stress that for calculation of the
susceptibility the external field must be weak,
$f\sqrt{N} \ll E_s-E_0 \sim 1/\sqrt{N}$. If this criterion is violated,
the ground state energy shift is proportional to $\propto f\sqrt{N}$
and the second derivative (\ref{chi1}) vanishes.

It is more likely that the bare singlet mass $m_s^2=-|m_s^2|$
is small, but non-zero. The dimer series expansions
\cite{Kotov} clearly show two nonzero dimer order parameters in the SL phase.
To avoid misunderstanding
we stress that we do not claim that one can derive a translationally
invariant ground state starting from the dimer basis. This is certainly
impossible. However if the true state is a translationally invariant
one then the dimer series must generate zero dimer order parameters
as an indication that there is something wrong with the dimer state.
This does not happen.
In principle it is certainly possible that convergence of series is so peculiar
that the standard  Pade approximant method based on a limited number of
terms does not give the right answer, but this is an unlikely event
and there no examples of such behavior.
In the case of the dimer ground state the static susceptibility of  
a finite cluster  with respect to the dimerization field (\ref{FD}) must 
scale  as
\begin{equation}
\label{chidd}
\chi = {1\over{2f^2}}{{|<0|F_d|0>|^2}\over{\Delta E}} \propto
{1\over{f^2}}{{(f N)^2}\over{\exp{(-aN)}}} \propto  N^2 e^{aN}.
\end{equation}
For the calculation the external field must be very weak,
$fN \ll \Delta E \sim \exp (-aN)$. If this criterion is violated,
the ground state energy shift is proportional to $\propto f N$
and the second derivative (\ref{chi1}) vanishes.

In conclusion we have demonstrated that the critical dynamics of singlet and 
triplet excitations in the two-dimensional $J_1-J_2$-model are closely related.
The critical points for singlet and triplet excitations are very close or 
coincide. This is a new type of critical behavior and we argue that 
this is a generic property of strongly frustrated spin systems. 
We have formulated the effective field theory describing the critical
behavior. The analysis favors gapless or almost gapless singlet 
excitations in the 
magnetically disordered phase. The effective field  theory deduced
in our work supports the  two-dimensional  extension of the 
Lieb-Schultz-Mattis theorem conjectured some time ago.

This work forms part of a research project supported by a grant from the 
Australian Research Council. 
We thank S. Sachdev for communicating important references.
The computation has been performed on the AlphaServer SC 
 computer. We are grateful for the computing resources provided
 by the Australian Partnership for Advanced Computing (APAC)
National Facility.

.
\begin{figure}[h]
\vspace{-65pt}
\hspace{-35pt}
\epsfxsize=8.5cm
\centering\leavevmode\epsfbox{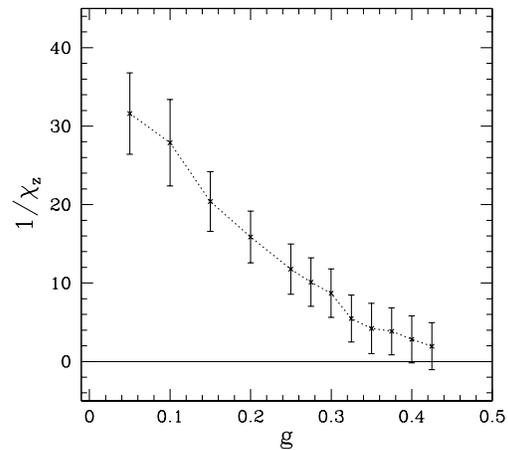}
\vspace{-60pt}
\caption{\it {The plot of $1/\chi_z$, where $\chi_z$ is the dimer 
susceptibility calculated in the N\'eel 
state using Ising series expansion. The value of 
$1/\chi_z$ vanishes at $g_{cs}=0.38\pm 0.04$.}}
\label{Fig1}
\end{figure}
\begin{figure}[h]
\vspace{-58pt}
\hspace{-35pt}
\epsfxsize=8.5cm
\centering\leavevmode\epsfbox{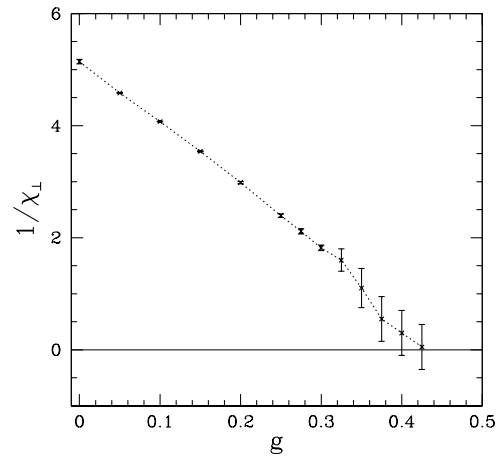}
\vspace{-60pt}
\caption{\it {The plot of $1/\chi_{\perp}$, where $\chi_{\perp}$ is the 
dimer susceptibility calculated 
in the N\'eel state using Ising series expansion. The value of 
$1/\chi_{\perp}$ vanishes at $g_{cs}=0.41\pm 0.03$.}}
\label{Fig2}
\end{figure}
\begin{figure}[h]
\vspace{-58pt}
\hspace{-35pt}
\epsfxsize=8.5cm
\centering\leavevmode\epsfbox{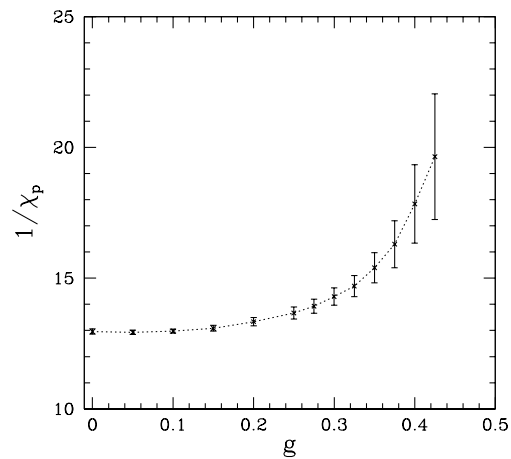}
\vspace{-60pt}
\caption{\it {The plot of $1/\chi_p$, where $\chi_p$ is the plaquette 
susceptibility calculated in the N\'eel 
state using Ising series expansion. The value of $1/\chi_p$ does not vanish,
so excitations with plaquette symmetry are not critical.}}
\label{Fig3}
\end{figure}

\end{document}